# Life Cycle Assessment of high rate algal ponds for wastewater treatment and resource recovery

Larissa Terumi Arashiro[a,b], Neus Montero[a], Ivet Ferrer[a], Francisco Gabriel Acién[c], Cintia Gómez[c], Marianna Garfí[a,*]

[a] GEMMA – Group of Environmental Engineering and Microbiology, Department of Civil and Environmental Engineering, Universitat Politècnica de Catalunya·BarcelonaTech, c/ Jordi Girona 1-3, Building D1, E-08034 Barcelona, Spain.

[b] Department of Industrial Biological Sciences, Ghent University, Graaf Karel de Goedelaan 5, 8500 Kortrijk, Belgium

[c] Department of Chemical Engineering, University of Almería, 04120 Almería, Spain

* Corresponding author:
Tel: +34 9340 16412
Fax: +34 93 4017357
Email: marianna.garfi@upc.edu

**Abstract**

The aim of this study was to assess the potential environmental impacts associated with high rate algal ponds (HRAP) systems for wastewater treatment and resource recovery in small communities. To this aim, a Life Cycle Assessment (LCA) was carried out evaluating two alternatives: i) a HRAP system for wastewater treatment where microalgal biomass is valorized for energy recovery (biogas production); ii) a HRAP system for wastewater treatment where microalgal biomass is reused for nutrients recovery (biofertilizer production). Additionally, both alternatives were compared to a typical small-sized activated sludge system. An economic assessment was also performed. The results showed that HRAP system coupled with biogas production appeared to be more environmentally friendly than HRAP system coupled with biofertilizer production in the climate change, ozone layer depletion, photochemical oxidant formation, and fossil depletion impact categories. Different climatic conditions have strongly influenced the results obtained in the eutrophication and metal depletion impact categories. In fact, the HRAP system located where warm temperatures and high solar radiation are predominant (HRAP system coupled with biofertilizer production) showed lower impact in those categories. Additionally, the characteristics (e.g. nutrients and heavy metals concentration) of microalgal biomass recovered from wastewater appeared to be crucial when assessing the potential environmental impacts in the terrestrial acidification, particulate matter formation and toxicity impact categories. In terms of costs, HRAP systems seemed to be more economically feasible when combined with biofertilizer production instead of biogas. On the whole, implementing HRAPs instead of activated sludge systems might increase sustainability and cost-effectiveness of wastewater treatment in small communities, especially if implemented in warm climate regions and coupled with biofertilizer production.

# 1. Introduction

High rate algal ponds (HRAPs) for wastewater treatment were introduced around 50 years ago and used since then not only to grow microalgae biomass but also to treat a wide variety of municipal and industrial wastewaters (Cragg et al., 2014; Oswald and Golueke, 1960). These systems are shallow, paddlewheel mixed, raceway ponds where microalgae assimilate nutrients and produce oxygen, which is used by heterotrophic bacteria to oxidise organic matter improving water quality (Craggs et al., 2014; Park et al., 2011). Since mechanical aeration is not required, energy consumption in these systems is much lower compared to a conventional wastewater treatment plant (e.g. activated sludge system) (around 0.02 kWh $m^{-3}$ of water vs. 1 kWh $m^{-3}$ of water, respectively) (Garfí et al., 2017; Passos et al., 2017). Moreover, HRAPs are less expensive and require little maintenance compared to conventional systems (Cragg et al., 2014; Garfí et al., 2017; Molinos-Senante et al., 2014). Due to their low cost and low energy consumption, HRAP systems could have a wide range of applications in Mediterranean regions, which present suitable climatic conditions for microalgae growth (e.g. high solar radiation). However, to achieve a satisfactory performance, large land area is required compared to conventional systems (around 6 $m^2$ p.e.$^{-1}$ vs. 0.5 $m^2$ p.e.$^{-1}$ for HRAP and activated sludge systems, respectively), making them more suitable for small communities (up to 10,000 p.e.).

Nowadays, there is an important need to shift the paradigm from wastewater treatment to resource recovery to alleviate negative effects associated with human activities, such as pollution of water bodies, greenhouse gas (GHG) emissions and scarcity of mineral resources. In this context, microalgae grown in HRAPs can be harvested and reused to produce biofuels or other non-food bioproducts. In particular, intensive research has been developed during the last years to investigate the potential of microalgae to produce biofuels such as biogas. Indeed, the biogas produced from microalgal biomass was found to contain high energy value, making microalgae anaerobic digestion an attractive alternative for biofuel production (Chew et al., 2017; Jankowska et al., 2017; Montingelli et al., 2015; Uggetti et al., 2017). On the other hand, microalgae also offer the potential to recover nutrients from wastewater and, subsequently, to be applied as a sustainable fertilizer. During the last decade, this alternative has been described by several authors, considering the fact that microalgae contain high amounts of proteins rich in essential amino acids, as well as phytohormones that stimulate plant growth (Coppens et al., 2016; García-Gonzalez and Sommerfeld, 2016; Jäger et al., 2010; Uysal et al., 2015).

Recent studies have employed the Life Cycle Assessment (LCA) methodology to assess the environmental impact of HRAP systems for wastewater treatment. They demonstrated that HRAPs might help to reduce environmental impacts and costs associated with wastewater treatment compared to conventional systems (e.g. activated sludge system), especially in small communities (Garfí et al., 2017; Maga, 2016). These studies also highlighted that the LCA methodology is an appropriate tool to support early-stage research and development of novel technologies and processes (Fang et al., 2016; Garfí et al., 2017). Indeed, LCA methodology takes into account and quantifies all environmental exchanges (i.e. resources, energy, emissions, waste) occurring during all stages of the technology life cycle (Ferreira et al., 2014; Ferreira et al., 2017; ISO, 2000).

Nevertheless, to the best of the authors' knowledge, there are no studies assessing the environmental impacts of HRAP system for wastewater treatment considering different configurations for resource and energy recovery.

The objective of this work was to evaluate the potential environmental impacts associated with HRAP systems for wastewater treatment taking into account two resource recovery strategies. To this aim a LCA was carried out comparing the following alternatives: (i) a HRAP system for wastewater treatment where microalgal biomass is valorised for energy recovery (biogas production); (ii) a HRAP system for wastewater treatment where microalgal biomass is reused for nutrients recovery (biofertilizer production). For the sake of comparison, both scenarios were compared to a typical small-sized activated sludge system. Additionally, an economic evaluation was addressed in order to assess the feasibility of the HRAP alternatives based on the costs and benefits related to each of them.

This paper is organized as follows: Section 2 describes the wastewater treatment systems, as well as the methodology used for the LCA and the economic analysis; in Section 3 the results of the comparative LCA and the economic analysis are described; finally, in Section 4 the main conclusions are highlighted.

## 2. Material and Methods

### 2.1 Wastewater treatment systems description

The HRAP systems were hypothetical wastewater treatment plants based on extrapolation from lab-scale and pilot-scale studies (up to 100 $m^2$). The systems were designed to serve a population equivalent of 10,000 p.e. and treat a flow rate of 1,950 $m^3$ $d^{-1}$. The HRAP system coupled with biogas production was considered to be implemented in Catalonia (Barcelona, Spain), where the mean temperature and global solar radiation are 15.5°C and 4.56 kWh/$m^2$d, respectively (AEMET, 2017). For this case study, the design parameters were calculated taking into account the experimental results obtained in lab-scale and pilot systems (up to 5 m2) located at the Universitat Politècnica de Catalunya-BarcelonaTech (UPC) (Barcelona, Spain) (García et al., 2000; García et al., 2006; Gutiérrez et al., 2016; Passos and Ferrer, 2014, Solé-Bundó et al., 2015; Solé-Bundó et al., 2017). This system comprises a primary settler (Hydraulic Retention Time (HRT) = 2.5 h) followed by four HRAPs (Table 1). From these units, wastewater goes through a secondary settler (HRT = 3 h) where microalgal biomass is harvested and separated from wastewater. Treated water is then discharged into a surface water body. Part of the harvested microalgal biomass (2 and 10 % on a dry weight basis in summer and winter, respectively) is recycled in order to enhance spontaneous flocculation (bioflocculation) and increase microalgae harvesting efficiency (Gutiérrez et al., 2016). The remaining harvested biomass is thickened (HRT = 24 h), thermally pretreated (75 °C, 10 h) and co-digested with primary sludge (35 °C, 20 days). The biogas produced is then converted in a combined heat and power (CHP) unit, while the digestate is transported and reused in agriculture. In this context, the HRT of each HRAP has to be modified over the year (8, 6 and 4 days) in accordance with the weather conditions (i.e. solar radiation and temperature) in order to accomplish wastewater treatment and meet effluent quality requirements for discharge (García et al., 2000; Gutiérrez et al., 2016). For this reason, it was considered that during summer months (from May to July) only two HRAPs work in parallel (HRT = 4 days), whereas all of them are operated during winter months (from November to April) (HRT = 8 days). During the rest of the year (from August to October), the HRT is 6 days (3 HRAPs working in parallel).

The HRAP system coupled with biofertilizer production was considered to be implemented in Andalucía (Almeria, Spain), where the mean temperature and global solar radiation are 19.1°C and 5.29 kWh/m2d, respectively (AEMET, 2017). For this case study, the designed parameters were determined using the results obtained in a pilot system located at the Las Palmerillas Expertimental Station (Almeria, Spain) (100 m2) (Morales-Amaral et al., 2015a). This system consists of two HRAPs operating in parallel and followed by a settler (HRT = 3 h) where microalgal biomass is separated using an organic flocculant (Table 2). From this unit, treated wastewater is discharged into a surface water body, while harvested microalgae biomass is dewatered on-site using a centrifuge and later sold to a local company to produce a biofertilizer (NPK = 5-1-0.75). The biofertilizer produced from the dewatered biomass is then transported and reused in agriculture. In this case, due to the more favourable climatic conditions for microalgae growth compared to Catalonia, the HRT was the same over the year (HRT = 3 days). It has to be noted that, for the same reason, the microalgal biomass production is considerably higher in the system implemented in Andalucía with respect to the one located in Catalonia (3-26 g$_{TSS}$ m$_{-2}$ d$_{-1}$ vs. 15-30 g$_{TSS}$ m$_{-2}$ d$_{-1}$, respectively) (Gutiérrez et al., 2016; Morales-Amaral et al., 2015a).

For the sake of comparison, the potential environmental impacts of the HRAP systems were compared to those generated by a conventional small-sized wastewater treatment plant (10,000 p.e.). For that purpose, the design of a usual small-scale activated sludge system implemented in Spain was taken into account (Gallego et al., 2008; Garfí et al., 2017; Lorenzo-Toja et al., 2015). It comprises a primary settler, followed by an activated sludge reactor with extended aeration and a secondary settler (Table 3). Treated water is discharged into the environment and the sludge is conditioned, thickened, centrifuged on-site and then transported to an incineration facility.

Figure 1 shows the flow diagrams of the treatment alternatives. Table 1, 2 and 3 show the characteristics and design parameters of the HRAP and activated sludge systems.

**Please insert Figure 1**

**Please insert Table 1**

**Please insert Table 2**

**Please insert Table 3**

*2.2 Life Cycle Assessment*

The LCA was conducted following the ISO standards (ISO, 2000; ISO, 2006) in order to evaluate and quantify the potential environmental impact of the investigated scenarios. It consisted of four main stages: i) goal and scope definition, ii) inventory analysis, iii) impacts assessment and iv) interpretation of the results (ISO, 2006). The following sections describe the specific content of each phase.

*2.2.1 Goal and scope definition*

The goal of this study was to determine the potential environmental impact of HRAP systems for wastewater treatment and resource recovery. In particular, two configurations were compared:

    a) a HRAP system for wastewater treatment where microalgal biomass is valorised for energy recovery (biogas production) (Scenario 1);

    b) a HRAP system for wastewater treatment where microalgal biomass is reused for nutrients recovery (biofertilizer production) (Scenario 2).

Moreover, both scenarios were compared to a typical small-sized activated sludge system implemented in Spain (Scenario 3). The functional unit (FU) for this study was set as 1 m$_3$ of treated water, since the main function of the technologies proposed is to treat wastewater.

The cradle-to-grave boundaries included systems construction, operation and maintenance over a 20-years period (Garfí et al., 2017; Pérez-López et al., 2017; Rahman et al., 2016) (Figure 1). Input and output flows of materials (i.e. construction materials and chemicals) and energy resources (heat and electricity) were systematically studied for all scenarios. Direct GHG emissions and $NH_4^+$ volatilization associated with wastewater treatment were also included in the boundaries. As treated water is discharged into the environment, direct emissions to water were also taken into account. Regarding digestate and biofertilizer reuse in agriculture in Scenarios 1 and 2, transportation (20 km) (Hospido et al., 2004) and direct emissions to soil (heavy metals), as well as direct GHG emissions, were accounted for. In the case of the activated sludge system (Scenario 3), inputs and outputs associated with sludge disposal (i.e. incineration) were also included in the boundaries. An average distance of 30 km was considered for sludge transportation to incineration facilities, based on circumstances generally observed in our zone. The end-of-life of infrastructures and equipment were neglected, since the impact would be marginal compared to the overall impact.

Since the studied scenarios would generate by-products (i.e. biogas, biofertilizer), the system expansion method has been used following the ISO guidelines (Guinée, 2002; ISO, 2006). In this method, by-products are supposed to avoid the production of conventional products. Thus, the impact related to conventional products is withdrawn from the overall impact of the system (Collet et al., 2011; ISO, 2006; Sfez et al., 2015). In this study, the digestate and the biofertilizer produced in HRAP systems coupled with biogas and biofertilizer production (Scenarios 1 and 2, respectively) were considered as substitutes to chemical fertilizer. Moreover, the avoided burdens of using heat and electricity produced in Scenario 1 (HRAP systems coupled with biogas production), instead of heat from natural gas and electricity supplied through the grid, were also considered.

*2.2.2 Inventory analysis*

Inventory data for the investigated scenarios are summarized in Table 4, 5 and 6. In the case of HRAP systems coupled with biogas and biofertilizer production (Scenarios 1 and 2), inventory data regarding construction materials and operation were based on the detailed engineering designs performed in the frame of this study. Treated wastewater characteristics were estimated considering the removal efficiencies and experimental results obtained in the pilot systems implemented at the Universitat Politècnica de Catalunya-BarcelonaTech (UPC) (5 $m_2$) (Gutiérrez et al., 2016) and at the Las Palmerillas Experimental Station (100 $m_2$) (Morales-Amaral et al., 2015a) for Scenarios 1 and 2, respectively. $NH_4^+$ volatilization was estimated through nitrogen mass balance. $NH_3$ and $N_2O$ emissions due to the application of digestate and biofertilizer on agricultural land were calculated using emissions factors from the literature (Hospido et al., 2008; IPCC, 2006; Lundin et al., 2000). In this case, $CH_4$ emissions were not considered since anaerobic decompositions do not occur if liquid fertilizer is used and the climate is predominantly dry (Hobson, 2003; Lundin et al., 2000). Heavy metals and nutrients (avoided Total Nitrogen (TN) and Total Phosphorous (TP)) content of the digestate and biofertilizer were gathered from experimental results obtained in the above-mentioned pilot systems (Morales-Amaral et al., 2015a; Solé-Bundó, et al., 2017). In order to estimate electricity and heat production from biogas cogeneration in Scenario 1 (HRAP systems coupled with biogas production), biogas production obtained in lab-scale experiments was taken into account (Solé-Bundó et al., 2015; Passos et al., 2017).

As mentioned above, data regarding the typical small-sized activated sludge system implemented in Spain (Scenario 3) were gathered from the literature (Gallego et al., 2008; Garfí et al., 2017; Lorenzo-Toja et al., 2015).

Background data (i.e. data of construction materials, chemicals, energy production, avoided fertilizer, transportation and sludge incineration process) were obtained from the *Ecoinvent 3.1* database (Moreno-Ruiz et al., 2014; Weidema et al., 2013). The Spanish electricity mix was used for all electricity requirements (Red Eléctrica Española, 2016).

**Please insert Table 4**

**Please insert Table 5**

**Please insert Table 6**

*2.2.3 Impact assessment*

The LCA was performed using the software *SimaPro® 8* (Pre-sustainability, 2014). Potential environmental impacts were calculated by the ReCiPe midpoint method (hierarchist approach) (Goedkoop et al., 2009). In this study, characterisation phase was performed considering the following impact categories: Climate Change, Ozone Depletion, Terrestrial Acidification, Freshwater Eutrophication, Marine Eutrophication, Photochemical Oxidant Formation, Particulate Matter Formation, Metal Depletion, Fossil Depletion, Human Toxicity and Terrestrial Ecotoxicity. These impact categories were selected according to the most relevant environmental issues related to wastewater treatment and used in previous LCA studies (Corominas et al., 2013; Fang et al., 2016;

Gallego et al., 2008; Garfí et al., 2017; Hospido et al., 2008). Normalisation was carried out in order to compare all the environmental impacts at the same scale. This provides information on the relative significance of the indicator results, allowing a fair comparison between the impacts estimated for each scenario (ISO, 2006). In this study, the European normalisation factors have been used (Europe ReCiPe H) (Goedkoop et al., 2009).

*2.3. Sensitivity analysis*

In order to evaluate the influence of the most relevant assumptions have on the results, a sensitivity analysis was performed considering the following parameters: $NH_3$ emissions due to the application of digestate and biofertilizer on agricultural land (Scenario 1 and 2); $N_2O$ emissions due to the application of digestate and biofertilizer on agricultural land (Scenario 1 and 2); digestate and biofertilizer transportation distance (Scenario 1 and 2). A variation of ± 10% was considered for all parameters and the sensitivity coefficient was calculated using Eq. (1) (Dixon et al., 2003):

$$\text{Sensitivity Coefficient (S)} = \frac{(\text{Output}_{high} - \text{Output}_{low})/\text{Output}_{default}}{(\text{Input}_{high} - \text{Input}_{low})/\text{Input}_{default}} \quad (1)$$

where Input is the value of the input variable (e.g. $NH_3$ and $N_2O$ emissions) and Output is the value of the environmental indicator (e.g. Climate Change).

*2.4 Seasonality*

Annual averages of potential environmental impacts from HRAPs scenarios (Scenario 1 and 2) were compared to those obtained considering the microalgal biomass production achieved in summer and winter months (highest and lowest production, respectively; Table 1 and 2) to assess their fluctuations over the year. In particular, the microalgal biomass production considered for Scenario 1 (HRAP systems coupled with biogas production) was 5 and 25 $g_{TSS}$ $m^{-2}$ $d^{-1}$ for winter and summer months, respectively. On the other hand, for Scenario 2 (HRAP systems coupled with biofertilizer production) a microalgal biomass production of 15 and 30 $g_{TSS}$ $m^{-2}$ $d^{-1}$ was considered for winter and summer months, respectively.

## *2.5 Economic assessment*

The economic assessment was performed comparing the capital cost and the operation and maintenance cost of Scenarios 1 and 2 (HRAP systems coupled with biogas and biofertilizer production, respectively). The capital cost included the cost for earthmoving and construction materials purchase. On the other hand, operation and maintenance cost comprised costs associated with energy (electricity and heat) consumption and chemicals purchase. In both scenarios, prices were provided by local companies. For Scenario 1 (HRAP systems coupled with biogas production), the surplus electricity generated from biogas cogeneration was supposed to be sold back to the grid. Thus, the price of electricity sold to the grid was withdrawn from the overall operational and maintenance cost of the system. For Scenario 2 (HRAP systems coupled with biofertilizer production), the dewatered microalgae biomass is sold to a local company (BIORIZON BIOTECH S.L.) to produce the biofertilizer (Romero-García et al., 2012). Therefore, its price was withdrawn from the overall operational and maintenance cost of the system. Other costs (e.g. labour costs, transportation) were assumed to be similar in both scenarios and, thus, were not included in the analysis.

## 3. Results and Discussion

### *3.1 Life Cycle Assessment*

#### *3.1.1 Characterization*

The potential environmental impacts associated with each alternative are shown in Figure 2. Comparing HRAP scenarios (Scenarios 1 and 2), the results show that Scenario 2 is the most environmentally friendly alternative in 7 out of 11 impact categories. As far as Climate Change, Ozone Depletion, Photochemical Oxidant Formation and Fossil Depletion Potentials are concerned, the potential environmental impact of Scenario 1 was lower than Scenario 2. This was mainly due to the offset energy generated from biogas cogeneration and the avoided fertilizer (Figure 2). In particular, the electricity generated by biogas cogeneration (avoided electricity) was around 9 times higher than that consumed for system operation in Scenario 1 (Table 4). It means that the surplus electricity could be sold to the grid. This is in accordance with previous studies that observed that, in a HRAP system for wastewater treatment, the energy balance is always positive when microalgal biomass is co-digested with primary sludge and the biogas is used to cogenerate electricity and heat (Passos et al., 2017). Moreover, it has to be noticed that the contribution of the avoided fertilizer to the overall impact was higher in Scenario 1 than Scenario 2 (Figure 2), since TN avoided was higher in the former compared to the latter (25.9 vs. 5.77 g $m^{-3}$ of water; Table 4 and 5). This can be explained by the fact that, despite TN content was higher in the biofertilizer (5 $g_{TN}$ $kg_{biofertilizer}^{-1}$) than in the digestate (1.89 $g_{TN}$ $kg_{digestate}^{-1}$), a lower amount of biofertilizer is produced in Scenario 2 (1.15 $kg_{biofertilizer}$ $m^{-3}$ of water) compared to Scenario 1 (13.7 $kg_{digestate}$ $m^{-3}$ of water). Indeed, the total solids (TS) content of the microalgal biomass obtained in Scenario 1 (2% TS) is lower compared to Scenario 2 (20%TS) due to its dewatering step (i.e. centrifugation). Nevertheless, it has to be mentioned that the biofertilizer is a higher quality product compared to the digestate, since it contains high amounts of proteins rich in essential amino acids, as well as phytohormones that stimulate plant growth and improve soil quality (Coppens et al., 2016; García-Gonzalez and Sommerfeld, 2016; Jäger et al., 2010; Uysal et al., 2015). However, these benefits were not taken into account in this study. Regarding Terrestrial Acidification and Particulate Matter Formation Potentials, Scenario 2 showed lower risks to endanger the environment because this configuration causes fewer emissions to air (i.e. $NH_3$ emissions) derived from biofertilizer application to agricultural soil compared to digestate from Scenario 1 (Table 4 and 5). With regards to Freshwater and Marine Eutrophication Potentials, Scenario 1 showed higher environmental impacts compared to Scenario 2. It is explained by the quality of treated effluent (i.e. lower TN and TP removal efficiencies in Scenario 1 than in Scenario 2; Table 4 and 5). The reason for this difference could be primarily due to the distinct climatic conditions, since the average temperature and global solar radiation in Catalonia (Scenario 1), as previously mentioned, are lower than in Andalucía (Scenario 2). Indeed, previous studies reported that nutrient removal efficiencies are improved with higher temperature and solar radiation (Craggs et al., 2012; Mehrabadi et al., 2016). Concerning Metal Depletion Potential, Scenario 1 would impair abiotic resources more likely than Scenario 2. Since Metal Depletion Potential is mainly influenced by construction materials, the lower environmental performance of Scenario 1 is owing to the larger surface area required for its implementation compared to Scenario 2 (4 $m^2$ p.e.$^{-1}$ vs. 3 $m^2$ p.e.$^{-1}$, respectively). As mentioned above, in the system implemented in Catalonia (Scenario 1), a higher HRT is needed (especially during winter months) compared to that implemented in Andalucía (Scenario 2) in order to obtain a effluent quality suitable for discharge (García et al., 2000; Gutiérrez et al., 2016, Morales-Amaral et al. 2015a; Morales-Amaral et al. 2015b). The influence of the geographical location on the performance of HRAPs was also addressed in previous studies, in which the use of this technology is not encouraged in northern regions, where the climatic conditions are not favourable to promote efficient wastewater treatment and biomass productivity (Grönlund and Fröling, 2014; Pérez-López et al., 2017). According to this, it is noteworthy to mention that, since in this study the two HRAP systems (Scenarios 1 and 2) were assumed to be implemented in locations with distinct climatic conditions, it is not possible to define the best biomass valorisation strategy (i.e. biogas vs. biofertilizer production). In fact, HRAP systems operating under similar conditions should be considered in order to enable a better comparison. In regard to Human toxicity and Terrestrial Ecotoxicity Potentials, Scenario 1 showed higher environmental impacts compared to Scenario 2 due to the higher concentration of heavy metals in the digestate than in the biofertilizer (Table 4 and 5).

According to the results presented in Figure 2, Scenarios 1 and 2 showed lower environmental impacts in 6 out of 11 impact categories (i.e. Climate Change, Ozone Depletion, Freshwater and Marine Eutrophication, Photochemical Oxidant Formation, Fossil Depletion) compared to Scenario 3. This was primarily due to the lower energy consumption needed for system operation in HRAP scenarios (Scenario 1 and 2) than in the activated sludge system (Scenario 3) (Table 4, 5 and 6). On the other hand, HRAP scenarios (Scenario 1 and 2) showed lower environmental performance in Metal Depletion category (Figure 2), since a higher amount of construction materials are needed for their implementation compared to the activated sludge system (Scenario 3). Indeed, even if HRAP systems have low raw materials requirements for their operation, a large amount of raw materials is needed for their construction. This fact could make HRAP systems less favourable than conventional technologies (e.g. activated sludge systems) in the abiotic resources depletion impact categories. Nevertheless, this drawback can be overcome by implementing HRAP systems in smaller agglomerations than that considered in this study (e.g. around 2,000 p.e.) (Garfí et al., 2017). As far as Terrestrial Acidification, Particulate Matter Formation, Human Toxicity and Terrestrial Ecotoxicity Potentials are concerned, the potential environmental impacts of HRAPs scenarios (Scenario 1 and 2) were higher than that caused by the activated sludge system (Scenario 3). It was mainly due to the $NH_3$ air emissions derived from $NH_4^+$ volatilization in HRAPs and to the heavy metals content in the digestate/biofertilizer (emissions to soil). The results are consistent with previous studies that reported increased toxicity in a comparative LCA by integrating a sidestream process into a conventional wastewater treatment facility where microalgae are cultivated, harvested and then used for fertigation (Fang et al., 2016). Furthermore, it was observed that the higher impacts on terrestrial environments are unavoidable in cases where sludge and nutrients from wastewater are recycled and reused in agriculture (Tangsubkul et al., 2005). In order to address this issue, improved technologies to separate better heavy metals from recycled sludge should be encouraged (Tangsubkul et al., 2005). In regard to Freshwater Eutrophication Potential, the activated sludge system (Scenario 3) showed higher potential environmental impact compared to Scenario 2, but lower impact than Scenario 1. This was because of the higher outlet Phosphorous concentration in Scenario 1 compared to the other scenarios, which might be related to the lower nutrients removal efficiency caused by less favourable climatic conditions. Previous studies observed that eutrophication and toxicity impact categories were mainly affected by water discharge emissions and sludge management, indicating that the best alternatives seem to be the ones that provide lower nutrients and heavy metals emissions (Corominas et al., 2013). This corroborates with the results obtained with this study, where the configuration with higher nutrients concentration in the effluent and higher levels of heavy metals in the recycled biomass (Scenario 1) showed higher impacts in those categories.

On the whole, HRAP systems coupled with biogas and biofertilizer production (Scenario 1 and 2) showed similar environmental performance if compared to the activated sludge system (Scenario 3). In particular, HRAPs environmental performance is better than the conventional system in the climate change, ozone layer depletion, photochemical oxidant formation, and fossil depletion impact categories. It was in accordance with previous studies, which stated that, compared to a typical medium-sized conventional wastewater treatment plant, a HRAP system coupled with biogas production could offer clear benefits with regard to the protection of climate, protection of fossil resources and ozone depletion (Maga, 2016). In order to reduce the environmental impacts of HRAP systems for wastewater treatment and resource recovery, the following improvements should be addressed and further assessed: i) reducing $NH_4^+$ volatilization in HRAPs by controlling the pH through $CO_2$ injection; ii) ensuring higher nutrients removal efficiencies by selecting a favourable geographical location to implement the HRAP systems; iii) studying improved technologies to separate heavy metals from recycled microalgal biomass; iv) improving HRAP design in order to decrease the amount of construction materials used (e.g. excavation instead of concrete structure).

**Please insert Figure 2**

*3.1.2 Normalization*

The normalised results show that Freshwater Eutrophication, Marine Eutrophication, Terrestrial Acidification and Human Toxicity Potentials are the most significant impact categories for all the scenarios considered (Figure 3). These results are in accordance with previous LCAs on wastewater treatment (Fang et al., 2016; Gallego et al, 2008; Hospido et al., 2004). In these impact categories, Scenario 2 showed to be the most environmentally friendly alternative.

**Please insert Figure 3**

*3.2 Sensitivity analysis*

The results of the sensitivity analysis are shown in Table 7, where the most sensitive inventory components are indicated by bold type.

The results showed that Terrestrial Acidification and Particulate Matter Formation Potentials are somewhat sensitive to $NH_3$ emissions due to the application of digestate on agricultural land in Scenario 1 (sensitivity coefficient around 0.3 for both environmental indicators). Indeed, a 10% increase of this parameter would increase these indicators by around 3%.

Similarly, Climate Change Potential showed to be somewhat sensitive to $N_2O$ emissions due to the application of digestate on agricultural land in Scenario 1 (sensitivity coefficient = 0.36). This means that a 10% increase in $N_2O$ direct emissions would increase this environmental indicator by 3.6%.

Moreover, Photochemical Oxidant Formation Potential showed to be sensitive to digestate transportation distance in Scenario 1 (sensitivity coefficient = 2.7). Indeed, a 10% increase in digestate transportation distance would increase this environmental indicator by 27%. The transport of the sludge to agricultural applications is not a fixed parameter, as it depends on specific needs. However, the sludge is usually applied in soil relatively close to the plant location (Pasqualino et al., 2009).

In conclusion, the results were found to be sensitive to digestate transportation distance in Scenario 1. Nevertheless, since it mainly affect only one of the less significant impact categories considered (i.e. Photochemical Oxidant Formation Potential), it can be concluded that the main findings of this study are not strongly dependent on the assumptions considered.

**Please insert Table 7**

*3.3 Seasonality*

The seasonal variation of the potential environmental impact for HRAPs scenarios (Scenario 1 and 2) are shown in Figure 4. The potential environmental impacts of Scenario 2 are fairly constant over the year. On the contrary, a strong seasonal variation was observed in Scenario 1. It was due to the fact that the microalgal biomass production range in Scenario 1 (5-25 $g_{TSS}$ $m^{-2}$ $d^{-1}$) is lower than Scenario 2 (15-30 $g_{TSS}$ $m^{-2}$ $d^{-1}$) and represents a high variation due to the seasonal fluctuations. It was in accordance with previous studies, which reported that meteorological conditions played a critical role in the LCA results of HRAPs for microalgal cultivation (Pérez-López et al., 2017). The authors highlighted that HRAPs are more suitable for locations where warm temperatures and high solar radiation are predominant (Pérez-López et al., 2017). Moreover, electricity and flocculants consumption, as well as water and biofertilizer characteristics, are fairly constant over the year in Scenario 2, while the biogas production and, consequently, the energy avoided, strongly depend on microalgal biomass production. These facts have a great influence on the environmental impacts seasonality in Scenario 1. As a result, Scenario 2 remained the most environmentally friendly alternative in 7 out of 11 impact categories compared to Scenario 1 over the year. Similarly, HRAPs scenarios (Scenario 1 and 2) still showed lower potential environmental impacts in 6 out of 11 impact categories compared to activated sludge system (Scenario 3) considering seasonal fluctuations.

**Please insert Figure 4**

*3.4 Economic assessment*

Results of the economic analysis are shown in Table 8. With respect to capital costs, Scenario 2 appeared as the less expensive alternative. It was due to its lower specific area requirement and, thus, lower amount of purchased materials, compared to Scenario 1 (3 vs. 4 $m_2$ p.e.$_{-1}$, respectively). Similar capital costs were found in previous studies which carried out an economic analysis of HRAPs for wastewater treatment without any resource recovery strategies (Garfí et al., 2017, Molinos-Senante et al., 2014). In fact, in this study the capital cost for ponds implementation was around 90% of the total capital cost of the overall systems (i.e. primary settler, ponds, secondary settler, digesters). Since the highest cost is due to ponds construction, implementing downstream units for resource recovery strategies (e.g. digester) in a HRAP system for wastewater treatment would slightly increase its capital costs. Regarding the operation costs, Scenario 2 showed to be the most expensive alternative, since this configuration requires higher expenses for energy and flocculant purchase. Nevertheless, if the price of the co-products (i.e. electricity sold back to the grid, microalgae biomass to produce the biofertilizer) that the wastewater treatment plant could sell out are considered, Scenario 2 would be the most cost-effective alternative (Table 8). The results of the economic assessment are consistent with previous studies, which indicated that recycling valuable compounds from microalgal biomass (such as nutrients and pigments) is likely to be more economically feasible than producing biogas from it, due to the higher added value of the final products (Ruiz et al., 2016; Vulsteke et al., 2017).

**Please insert Table 8**

## 4. Conclusions

In this study, the LCA methodology was a useful tool to identify the main environmental bottlenecks to scale-up high rate algal pond (HRAP) systems for wastewater treatment and resource recovery in small communities.

Results showed that HRAP system coupled with biogas production showed to be more environmentally friendly than HRAP system coupled with biofertilizer production in the climate change, ozone layer depletion, photochemical oxidant formation, and fossil depletion impact categories. Different climatic conditions have strongly influenced the results obtained in the eutrophication and metal depletion impact categories. In fact, the HRAP system located where warm temperatures and high solar radiation are predominant (HRAP system coupled with biofertilizer production) showed lower impact in those categories due to its higher nutrients removal efficiencies and lower hydraulic retention time (i.e. lower specific area requirement). The characteristics (e.g. total solids, nutrients and heavy metals concentration) of microalgal biomass recovered from wastewater appeared to be crucial when assessing the potential environmental impacts in the terrestrial acidification, particulate matter formation and toxicity impact categories.

Normalization identified Freshwater Eutrophication, Marine Eutrophication, Terrestrial Acidification and Human Toxicity as the most significant impact categories for all the scenarios considered. In these categories, HRAP system coupled with biofertilizer production and implemented in warm climate region showed to be the most environmentally friendly alternative.

Additionally, HRAP systems coupled with biogas and biofertilizer production showed lower potential environmental impacts compared to an activated sludge system in the climate change, ozone layer depletion, photochemical oxidant formation, and fossil depletion impact categories.

The environmental performance of HRAP technology for wastewater treatment and resource recovery in small communities might be improved by: i) reducing $NH_4^+$ volatilization in HRAPs by controlling the pH through $CO_2$ injection; ii) ensuring higher nutrients removal efficiencies by selecting a favourable geographical location to implement the HRAP systems; iii) studying improved technologies to separate heavy metals from recycled microalgal biomass; iv) improving HRAP design in order to decrease the amount of construction materials used.

In terms of costs, HRAP system coupled with biofertilizer production was the most cost-effective alternative, due to the higher added value of the biofertilizer compared to the energy obtained from biogas cogeneration.

In conclusion, HRAPs are sustainable and cost-effective technology for wastewater treatment in small communities, especially if implemented in warm climate regions and coupled with biofertilizer production. Their implementation and dissemination can help to support a shift towards resource recovery and a sustainable circular economy.

**Acknowledgements**

This research was funded by the Spanish Ministry of Economy and Competitiveness (FOTOBIOGAS CTQ2014-57293-C3-3-R; EDARSOL CTQ2014-57293-C3-1-R) and the European Union's Horizon 2020 research and innovation programme under the Marie Skłodowska-Curie grant agreement No 676070. This communication reflects only the authors' view and the Research Executive Agency of the EU is not responsible for any use that may be made of the information it contains.

**Table 1.** Characteristics and design parameters of the HRAP coupled with biogas production (Scenario 1)

| System characteristics | Unit | | | |
|---|---|---|---|---|
| Inlet BOD$_5$ concentration | $mg_{BOD}$ $L^{-1}$ | 300 | | |
| Outlet BOD$_5$ concentration | $mg_{BOD}$ $L^{-1}$ | <25 | | |
| Inlet TSS concentration | $mg_{TSS}$ $L^{-1}$ | 150 | | |
| Outlet TSS concentration | $mg_{TSS}$ $L^{-1}$ | <35 | | |
| Inlet Total Nitrogen | $mg_{TN}$ $L^{-1}$ | 39 | | |
| Outlet Total Nitrogen | $mg_{TN}$ $L^{-1}$ | 9.38 | | |
| Inlet Total Phosphorous | $mg_{TP}$ $L^{-1}$ | 5 | | |
| Outlet Total Phosphorous | $mg_{TP}$ $L^{-1}$ | 3.69 | | |
| Flow rate | $m^3$ $d^{-1}$ | 1,950 | | |
| Population equivalent | p.e. | 10,000 | | |
| Total surface area | $m^2$ | 40,000 | | |
| Specific area requirement | $m^2$ p.e.$^{-1}$ | 4 | | |
| **HRAPs Design parameters** | **Unit** | **Summer** | **Winter** | **Rest of the year** |
| OLR | $g_{BOD}$ $m^{-2}$ $d^{-1}$ | 10 | | |
| HRT | d | 4 | 8 | 6 |
| Number of ponds | - | 2 | 4 | 3 |
| Channel width | m | 12 | | |
| Channel length | m | 812.5 | | |
| Water depth | m | 0.4 | | |
| Microalgae biomass production | $g_{TSS}$ $m^{-2}$ $d^{-1}$ | 25.8 | 3.3 | 10.5 |
| Annual average microalgae biomass production | $g_{TSS}$ $m^{-2}$ $d^{-1}$ | 12 | | |

*Note: BOD: Biochemical oxygen demand; TSS: Total suspended solids; HRT: Hydraulic Retention Time; OLR: Organic Loading Rate. Summer: from May to July; winter: from November to April.*

**Table 2.** Characteristics and design parameters of the HRAP coupled with biofertilizer production (Scenario 2)

| System characteristics | Unit | | | |
|---|---|---|---|---|
| Inlet BOD$_5$ concentration | mg$_{BOD}$ L$^{-1}$ | 300 | | |
| Outlet BOD$_5$ concentration | mg$_{BOD}$ L$^{-1}$ | <25 | | |
| Inlet TSS concentration | mg$_{TSS}$ L$^{-1}$ | 200 | | |
| Outlet TSS concentration | mg$_{TSS}$ L$^{-1}$ | <35 | | |
| Inlet Total Nitrogen | mg$_{TN}$ L$^{-1}$ | 50 | | |
| Outlet Total Nitrogen | mg$_{TN}$ L$^{-1}$ | 2 | | |
| Inlet Total Phosphorous | mg$_{TP}$ L$^{-1}$ | 10 | | |
| Outlet Total Phosphorous | mg$_{TP}$ L$^{-1}$ | 1 | | |
| Flow rate | m$^3$ d$^{-1}$ | 1,950 | | |
| Population equivalent | p.e. | 10,000 | | |
| Total surface area | m$^2$ | 30,000 | | |
| Specific area requirement | m$^2$ p.e.$^{-1}$ | 3 | | |
| **HRAPs Design parameters** | Unit | Summer | Winter | Rest of the year |
| OLR | g$_{BOD}$ m$^{-2}$ d$^{-1}$ | 20 | | |
| HRT | d | 3 | | |
| Number of ponds | - | 2 | | |
| Channel width | m | 12 | | |
| Channel length | m | 1,219 | | |
| Water depth | m | 0.2 | | |
| Microalgae biomass production | g$_{TSS}$ m$^{-2}$ d$^{-1}$ | 30 | 15 | 25 |
| Annual average microalgae biomass production | g$_{TSS}$ m$^{-2}$ d$^{-1}$ | 23 | | |

*Note: BOD: Biochemical oxygen demand; TSS: Total suspended solids; HRT: Hydraulic Retention Time; OLR: Organic Loading Rate. Summer: from May to August; winter: from November to March*

**Table 3.** Characteristics and design parameters of the activated sludge system (Scenario 3)

| System characteristics | Unit | |
|---|---|---|
| Inlet BOD$_5$ concentration | $mg_{BOD}$ $L^{-1}$ | 300 |
| Outlet BOD$_5$ concentration | $mg_{BOD}$ $L^{-1}$ | <25 |
| Outlet TSS concentration | $mg_{TSS}$ $L^{-1}$ | <35 |
| Flow rate | $m^3$ $d^{-1}$ | 1,950 |
| Population equivalent | p.e. | 10,000 |
| Total surface area | $m^2$ | 900 |
| Specific area requirement | $m^2$ $p.e.^{-1}$ | 0.6 |
| **Design parameters** | Unit | |
| Primary settler HRT | h | 2.5 |
| Activated sludge reactor HRT | h | 6 |
| Secondary settler HRT | h | 2 |

*Note: BOD: Biochemical oxygen demand; TSS: Total suspended solids; HRT: Hydraulic Retention Time;*

*OLR: Organic Loading Rate.*

**Table 4.** Summary of the inventory for Scenario 1: HRAP system for wastewater treatment where microalgal biomass is valorised for energy recovery (biogas production). Values are referred to the functional unit (1 m$^3$ of water)

| Inputs | Scenario 1 | Units |
|---|---|---|
| **_Construction materials_** | | |
| _Primary settler_ | | |
| Concrete | 2.55E-06 | _m$^3$ m$^{-3}$_ |
| Steel | 2.04E-04 | _kg m$^{-3}$_ |
| _HRAPs_ | | |
| Concrete | 5.94E-04 | _m$^3$ m$^{-3}$_ |
| Steel | 4.76E-02 | _kg m$^{-3}$_ |
| _Secondary settler_ | | |
| Concrete | 1.29E-05 | _m$^3$ m$^{-3}$_ |
| Steel | 1.03E-03 | _kg m$^{-3}$_ |
| _Thickener_ | | |
| Concrete | 1.78E-07 | _m$^3$ m$^{-3}$_ |
| Steel | 1.42E-05 | _kg m$^{-3}$_ |
| _Thermal pretreatment_ | | |
| Concrete | 2.77E-07 | _m$^3$ m$^{-3}$_ |
| Steel | 2.22E-05 | _kg m$^{-3}$_ |
| _Digester_ | | |
| Concrete | 9.79E-06 | _m$^3$ m$^{-3}$_ |
| Steel | 7.83E-04 | _kg m$^{-3}$_ |
| **_Operation_** | | |
| _Energy consumption*_ | | |
| Primary settler | 4.41E-03 | _kWh m$^{-3}$_ |
| HRAPs | 1.13E-02 | _kWh m$^{-3}$_ |
| Secondary settler | 2.52E-03 | _kWh m$^{-3}$_ |
| Thermal pretreatment | 1.08E-04 | _kWh m$^{-3}$_ |
| Digester | 4.17E-02 | _kWh m$^{-3}$_ |
| Total energy consumption | 6.00E-02 | _kWh m$^{-3}$_ |
| **Outputs** | | |
| **_Emissions to water*_** | | |
| Total COD | 7.63E+01 | _g m$^{-3}$_ |
| TSS | 2.40E+01 | _g m$^{-3}$_ |
| TN | 9.38E+00 | _g m$^{-3}$_ |
| TP | 3.69E+00 | _g m$^{-3}$_ |
| **_Emissions to air*_** | | |
| _NH$_4^+$ volatilization in HRAPs_ | | |
| NH$_3$ | 3.80E+00 | _g m$^{-3}$_ |
| _Digestate application as fertilizer_ | | |
| NH$_3$ | 6.47E+00 | _g m$^{-3}$_ |
| N$_2$O | 2.59E-01 | *g m$^{-3}$* |
| **Emissions to soil*** | | |
| *Digestate application as fertilizer* | | |
| Cd | 3.53E-03 | *g m$^{-3}$* |
| Cu | 2.02E-01 | *g m$^{-3}$* |
| Pb | 9.08E-02 | *g m$^{-3}$* |
| Zn | 9.04E-01 | *g m$^{-3}$* |
| Ni | 4.15E-02 | *g m$^{-3}$* |
| Cr | 5.22E-02 | *g m$^{-3}$* |
| Hg (value <) | 4.52E-04 | *g m$^{-3}$* |
| **Avoided products*** | | |
| Electricity (from biogas cogeneration) | 5.40E-01 | *kWh m$^{-3}$* |
| Heat (from biogas cogeneration) | 8.49E-01 | *kWh m$^{-3}$* |
| N as Fertiliser (from digestate reuse) | 2.59E+01 | *g m$^{-3}$* |
| P as Fertiliser (from digestate reuse) | 1.31E+00 | *g m$^{-3}$* |

*\* Annual averages*

**Table 5.** Summary of the inventory for Scenario 2: HRAP system for wastewater treatment where microalgal biomass is reused for nutrients recovery (biofertilizer production). Values are referred to the functional unit (1 m$^3$ of water)

| Inputs | Scenario 2 | Units |
|---|---|---|
| **Construction materials** | | |
| *HRAPs* | | |
| Concrete | 4.32E-04 | *m$^3$ m$^{-3}$* |
| Steel | 3.45E-02 | *kg m$^{-3}$* |
| *Secondary settler* | | |
| Concrete | 1.29E-05 | *m$^3$ m$^{-3}$* |
| Steel | 1.03E-03 | *kg m$^{-3}$* |
| *Centrifuge* | | |
| Steel | 3.86E-05 | *kg m$^{-3}$* |
| **Operation** | | |
| *Energy consumption\** | | |
| HRAPs | 1.11E-02 | *kWh m$^{-3}$* |
| Secondary settler | 5.77E-03 | *kWh m$^{-3}$* |
| Centrifuge | 1.15E-02 | *kWh m$^{-3}$* |
| Biofertilizer production | 4.70E-02 | *kWh m$^{-3}$* |
| Total energy consumption | 7.54E-02 | *kWh m$^{-3}$* |
| *Chemicals\** | | |
| Organic flocculant | 1.00E+01 | *kg m$^{-3}$* |
| **Outputs** | | |
| *Emissions to water\** | | |
| Total COD | 1.00E+02 | *g m$^{-3}$* |
| TSS | 5.00E+01 | *g m$^{-3}$* |
| TN | 2.00E+00 | *g m$^{-3}$* |
| TP | 1.00E+00 | *g m$^{-3}$* |
| *Emissions to air\** | | |
| *NH$_{4+}$ volatilization in HRAPs* | | |
| NH$_3$ | 5.00E+00 | *g m$^{-3}$* |
| *Biofertilizer* | | |
| NH$_3$ | 1.44E+00 | *g m$^{-3}$* |
| N$_2$O | 5.77E-02 | *g m$^{-3}$* |
| *Emissions to soil\** | | |
| *Biofertilizer* | | |
| Cd | 3.46E-04 | *g m$^{-3}$* |
| Cu | 4.62E-02 | *g m$^{-3}$* |
| Pb | 2.31E-02 | *g m$^{-3}$* |
| Zn | 1.15E-02 | *g m$^{-3}$* |
| Ni | 1.15E-02 | *g m$^{-3}$* |
| Cr | 3.46E-02 | *g m-3* |
| Hg (value <) | 2.31E-04 | *g m-3* |
| **Avoided products*** | | |
| N as Fertiliser (from biofertilizer) | 5.77E+00 | *g m-3* |
| P as Fertiliser (from biofertilizer) | 1.20E+00 | *g m-3* |

*\* Annual averages*

**Table 6.** Summary of the inventory for Scenario 3: typical small-sized activated sludge system implemented in Spain. Values are referred to the functional unit (1 $m^3$ of water)

| Inputs | Scenario 3 | Units |
|---|---|---|
| ***Construction materials*** | | |
| Concrete | 1.65E-05 | $m^3\ m^{-3}$ |
| Steel | 1.32E-03 | $kg\ m^{-3}$ |
| ***Operation*** | | |
| *Energy consumption* | | |
| Electricity | 8.90E-01 | $kWh\ m^{-3}$ |
| *Chemicals* | | |
| Polyelectrolyte | 1.98E+00 | $g\ m^{-3}$ |
| Coagulant | 3.18E+00 | $g\ m^{-3}$ |
| **Outputs** | | |
| ***Emissions to water*** | | |
| Total COD | 1.25E+02 | $g\ m^{-3}$ |
| TSS | 3.50E+01 | $g\ m^{-3}$ |
| TN | 1.50E+01 | $g\ m^{-3}$ |
| TP | 2.00E+00 | $g\ m^{-3}$ |
| ***Emissions to air*** | | |
| $CO_2$ | 1.70E-01 | $g\ m^{-3}$ |
| $N_2O$ | 1.10E-01 | $g\ m^{-3}$ |
| ***Waste to further treatment*** | | |
| Sludge (incineration) | 1.24E+00 | $kg\ m^{-3}$ |

**Table 7**. Results of the sensitivity analysis for the considered parameters: NH₃ emissions due to the application of digestate and biofertilizer on agricultural land; N₂O emissions due to the application of digestate and biofertilizer on agricultural land; digestate and biofertilizer transportation distance.

| Impact categories | *Parameters* | | | | | |
|---|---|---|---|---|---|---|
| | *Scenario 1* | | | *Scenario 2* | | |
| | *NH₃ emissions* | *N₂O emissions* | *Digestate transportation* | *NH₃ emissions* | *N₂O emissions* | *Biofertilizer transportation* |
| Climate change | ±0.000 | ***±0.367*** | ±0.260 | ±0.000 | ±0.068 | ±0.015 |
| Ozone Depletion | ±0.000 | ±0.000 | ±0.204 | ±0.000 | ±0.000 | ±0.053 |
| Terrestrial acidification | ***±0.337*** | ±0.000 | ±0.008 | ±0.213 | ±0.000 | ±0.001 |
| Freshwater eutrophication | ±0.000 | ±0.000 | ±0.001 | ±0.000 | ±0.000 | ±0.000 |
| Marine eutrophication | ±0.058 | ±0.000 | ±0.001 | ±0.052 | ±0.000 | ±0.000 |
| Photochemical oxidant formation | ±0.000 | ±0.000 | ***±2.713*** | ±0.000 | ±0.000 | ±0.025 |
| Particulate matter formation | ***±0.327*** | ±0.000 | ±0.033 | ±0.179 | ±0.000 | ±0.003 |
| Metal depletion | ±0.000 | ±0.000 | ±0.019 | ±0.000 | ±0.000 | ±0.002 |
| Fossil depletion | ±0.000 | ±0.000 | ±0.153 | ±0.000 | ±0.000 | ±0.027 |
| Human toxicity | ±0.000 | ±0.000 | ±0.021 | ±0.000 | ±0.000 | ±0.011 |
| Terrestrial ecotoxicity | ±0.000 | ±0.000 | ±0.019 | ±0.000 | ±0.000 | ±0.011 |

*Note: Scenario 1: HRAP system for wastewater treatment where microalgal biomass is valorized for energy recovery (biogas production); Scenario 2: HRAP system for wastewater treatment where microalgal biomass is reused for nutrients recovery (biofertilizer production)*

**Table 8.** Results of the economic analysis for the HRAPs scenarios.

| | Unit | Scenario 1 | Scenario 2 |
|---|---|---|---|
| Capital cost | € p.e.$^{-1}$ | 192.55 | 139.34 |
| Operation and maintenance cost (energy and flocculant consumption) | € m$^{-3}$water | 0.007 | 0.02 |
| Price of electricity sold back to the grid | € m$^{-3}$water | 0.014 | - |
| Price of microalgal biomass sold to a company to produce the biofertilizer | € m$^{-3}$water | - | 8.08 |
| Profit (calculated considering operation cost only) | € m$^{-3}$water | 0.007 | 8.06 |

*Note: Scenario 1: HRAP system for wastewater treatment where microalgal biomass is valorised for energy recovery (biogas production); Scenario 2: HRAP system for wastewater treatment where microalgal biomass is reused for nutrients recovery (biofertilizer production)*

(a)

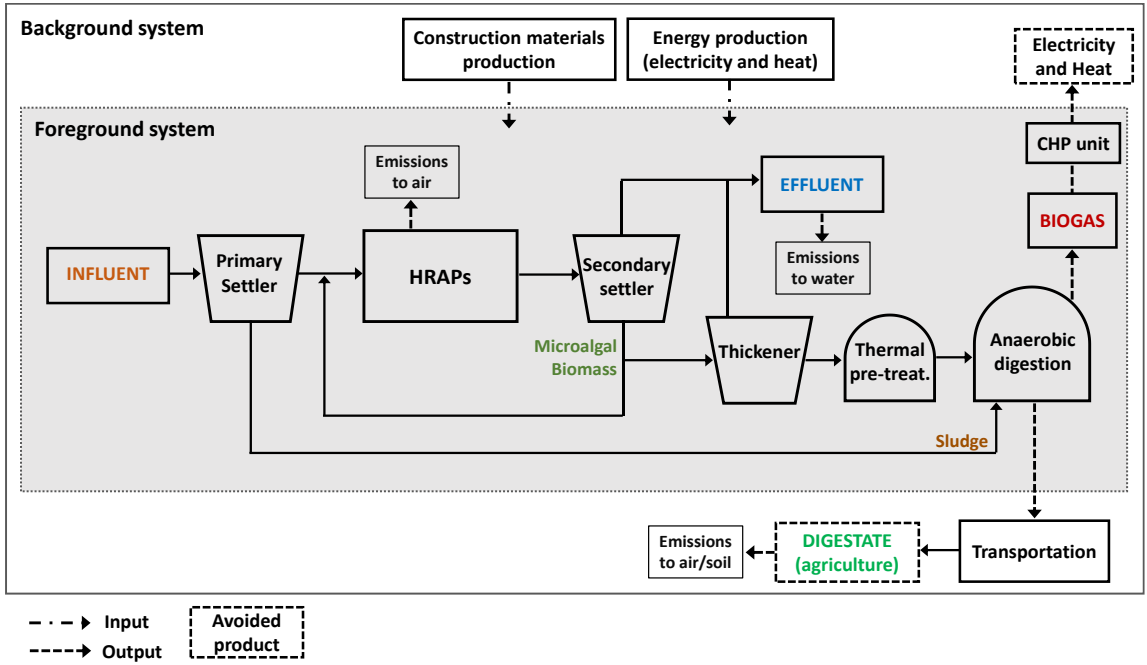

(b)

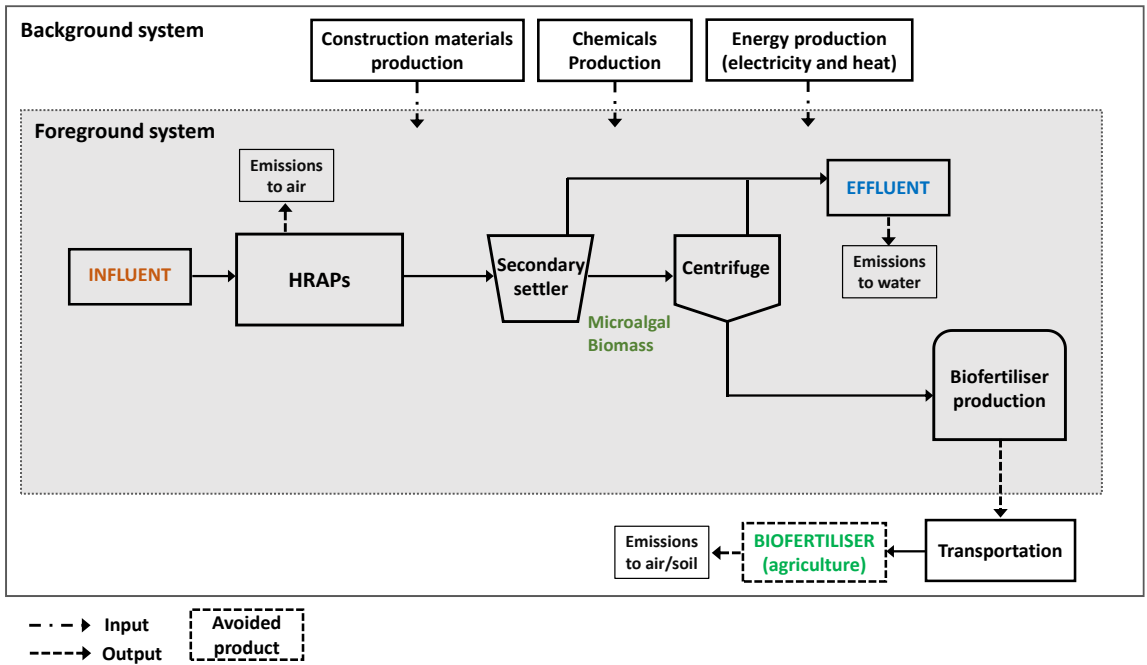

(c)

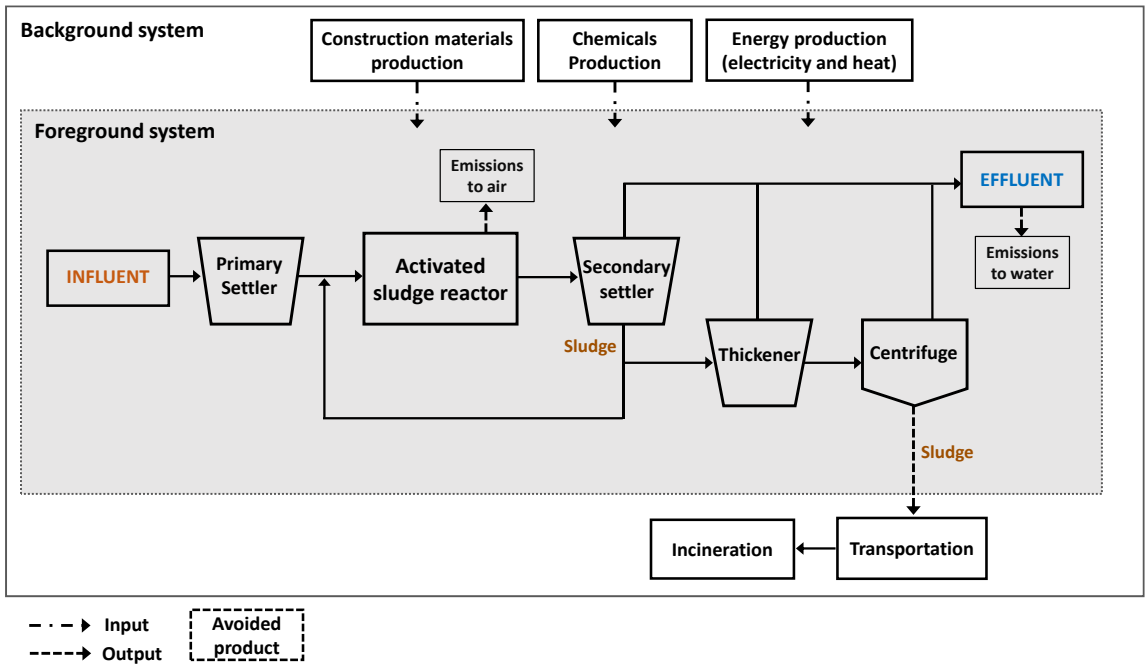

**Figure 1.** Flow diagrams and system boundaries of the wastewater treatment alternatives: a) HRAP system for wastewater treatment where microalgal biomass is valorised for energy recovery (biogas production) (Scenario 1); b) HRAP system for wastewater treatment where microalgal biomass is reused for nutrients recovery (biofertilizer production) (Scenario 2); c) activated sludge system (Scenario 3)

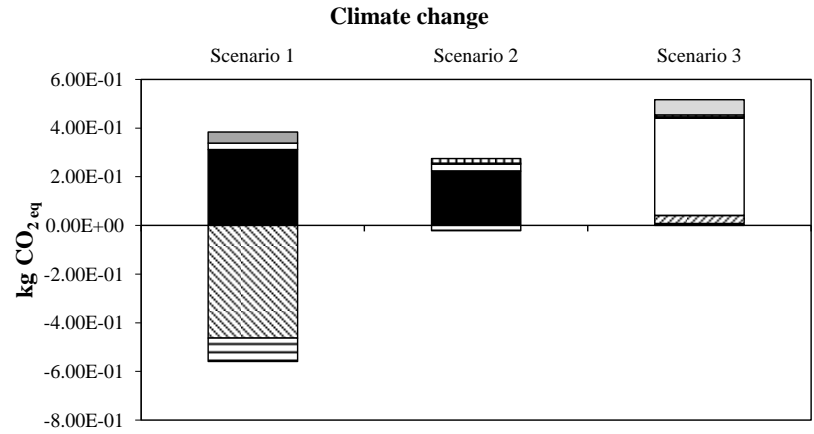
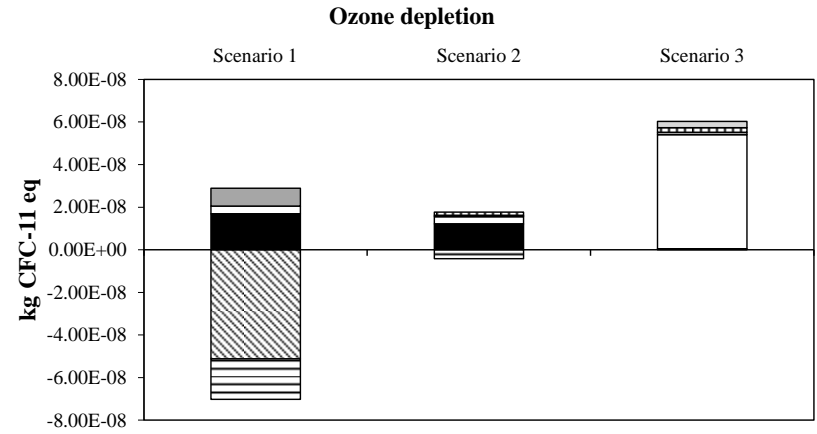
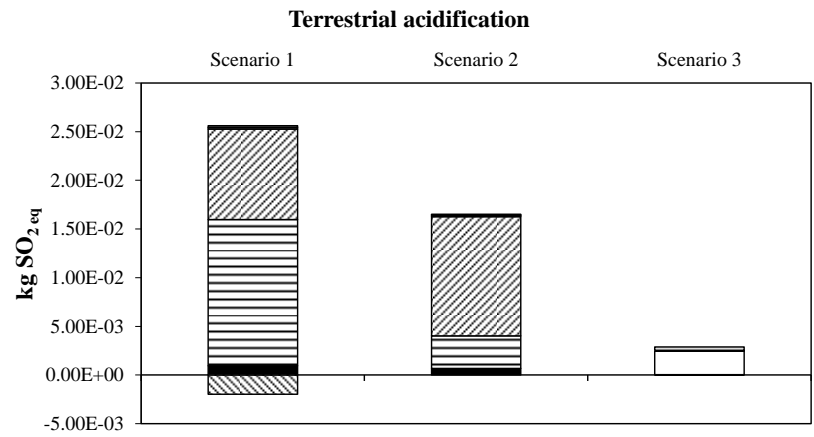
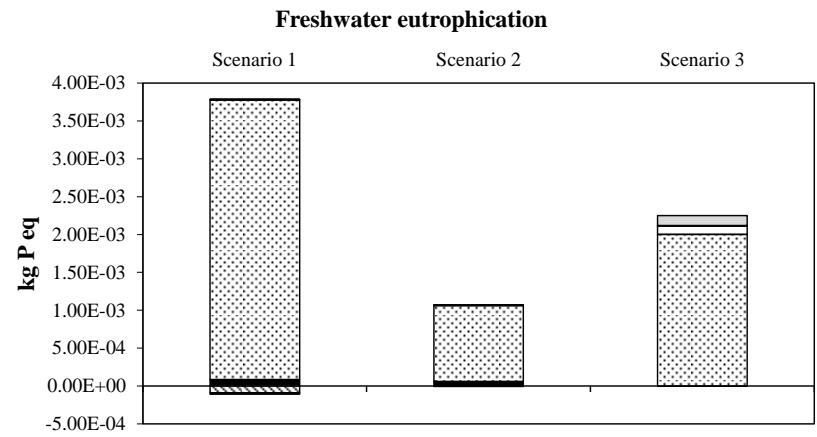

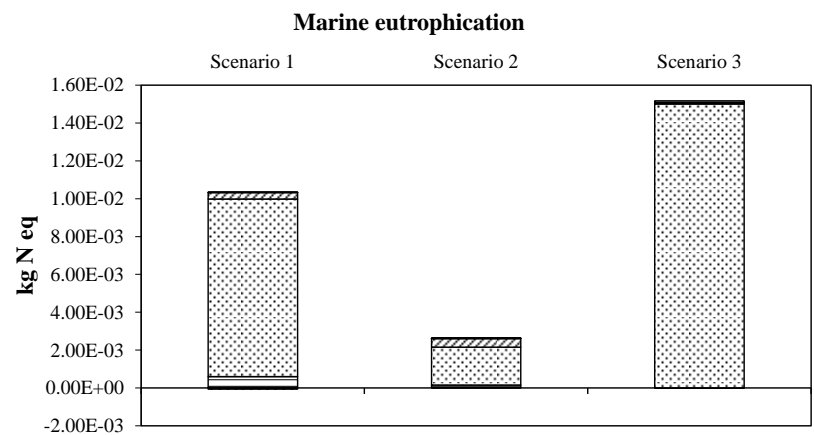
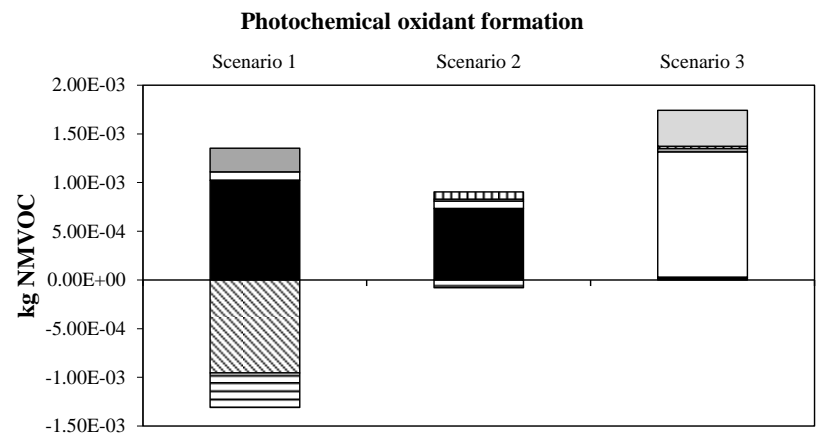
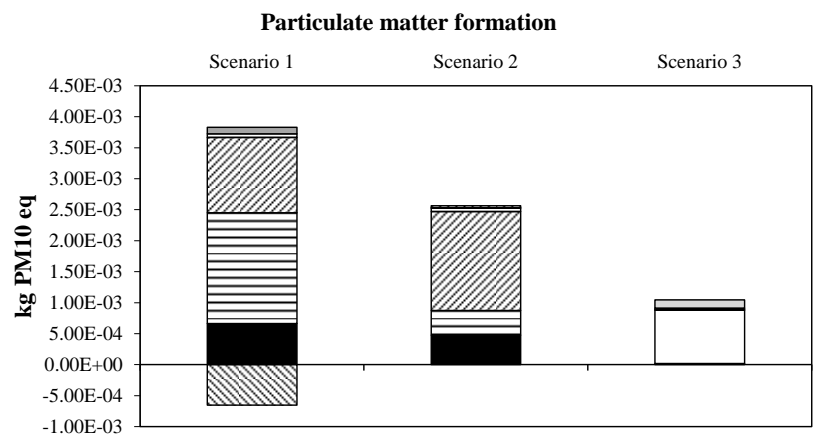
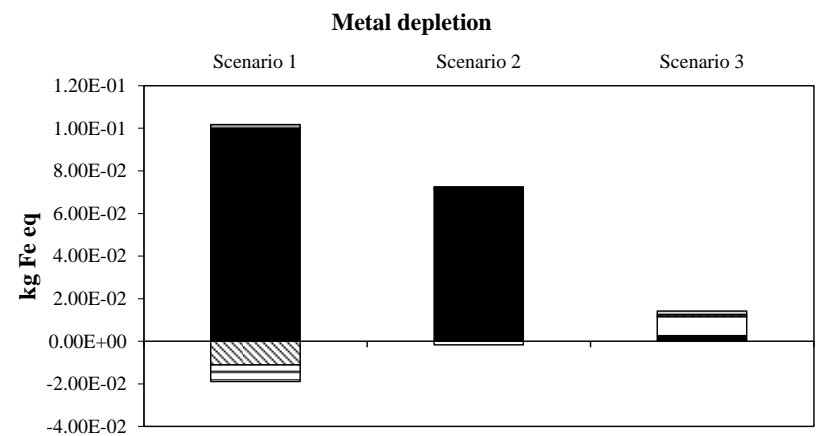

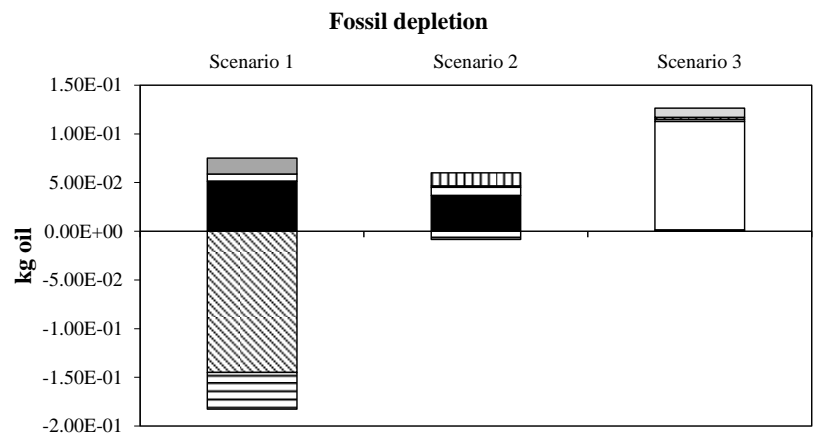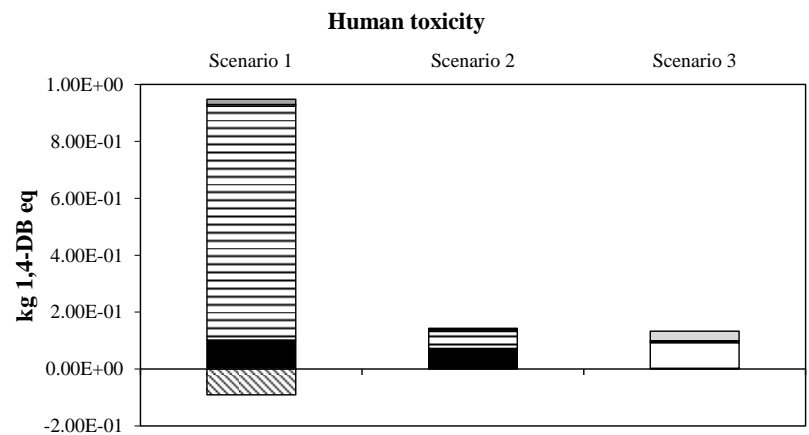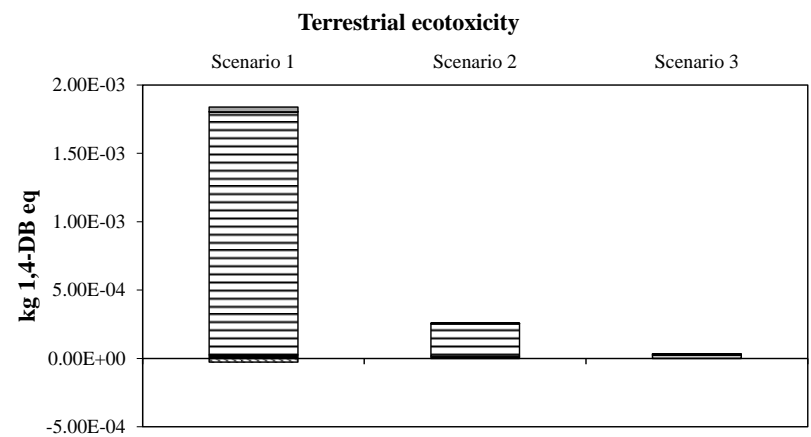

■ Construction materials
⊟ Digestate and biofertilizer application (including avoided fertilizer)
▨ Emissions to air (NH4+ volatilization in HRAP)
▥ Digestate, biofertilizer or sludge transportation
☐ Sludge disposal

⊠ Biogas cogeneration and avoided energy
⊠ Emissions to water
☐ Energy consumption
⊞ Chemicals

**Figure 2.** Potential environmental impacts for the three scenarios: a) HRAP system for wastewater treatment where microalgal biomass is valorised for energy recovery (biogas production) (Scenario 1); b) HRAP system for wastewater treatment where microalgal biomass is reused for nutrients recovery (biofertilizer production) (Scenario 2); c) activated sludge system (Scenario 3). Values are referred to the functional unit (1 m3 of water).

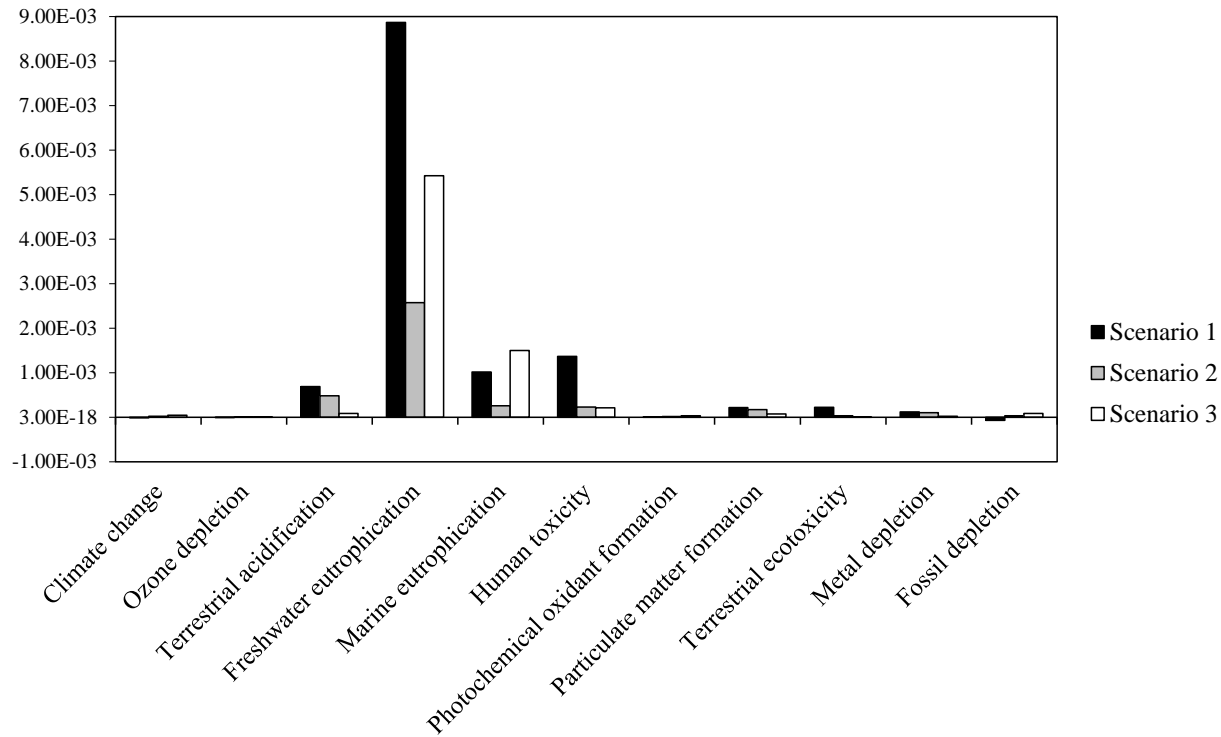

**Figure 3.** Normalised potential environmental impacts for the three scenarios: a) HRAP system for wastewater treatment where microalgal biomass is valorised for energy recovery (biogas production) (Scenario 1); b) HRAP system for wastewater treatment where microalgal biomass is reused for nutrients recovery (biofertiliser production) (Scenario 2); c) activated sludge system (Scenario 3).

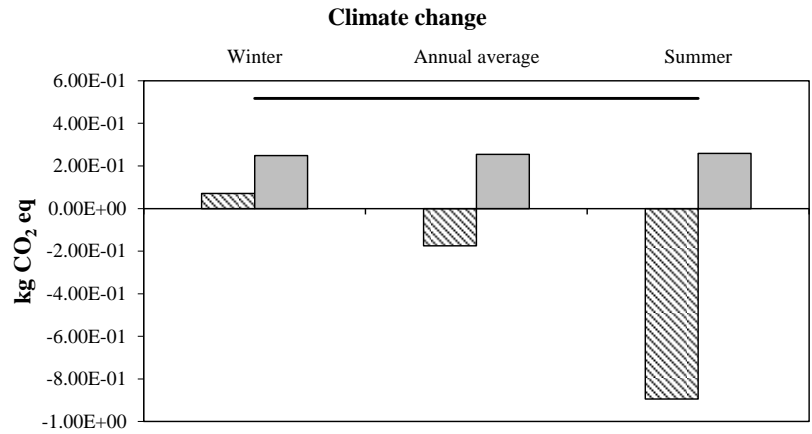
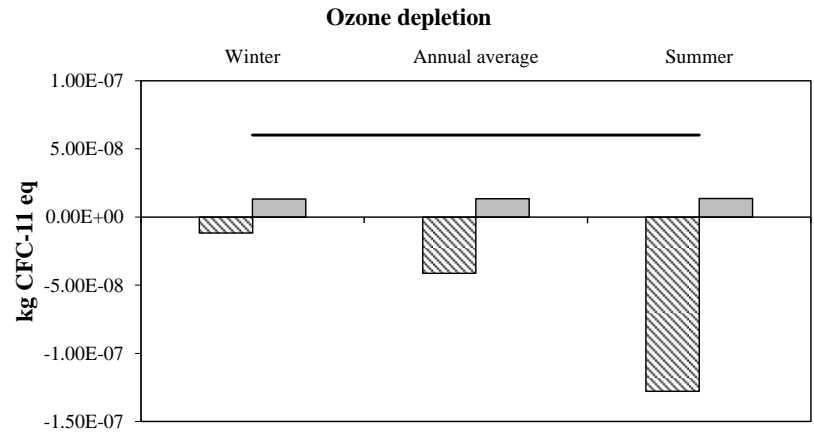
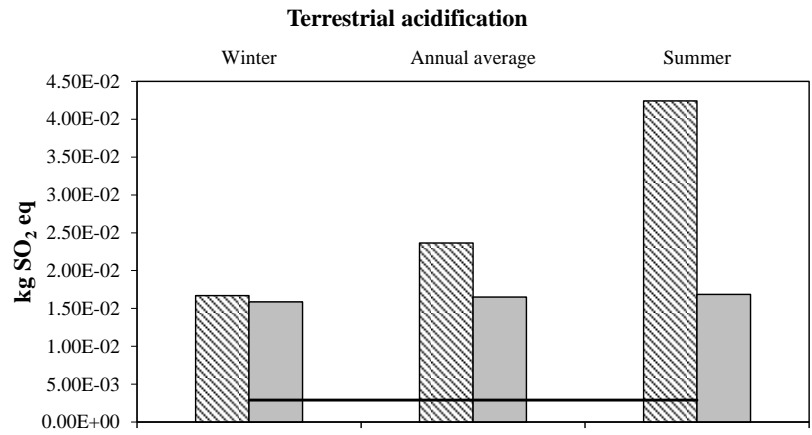
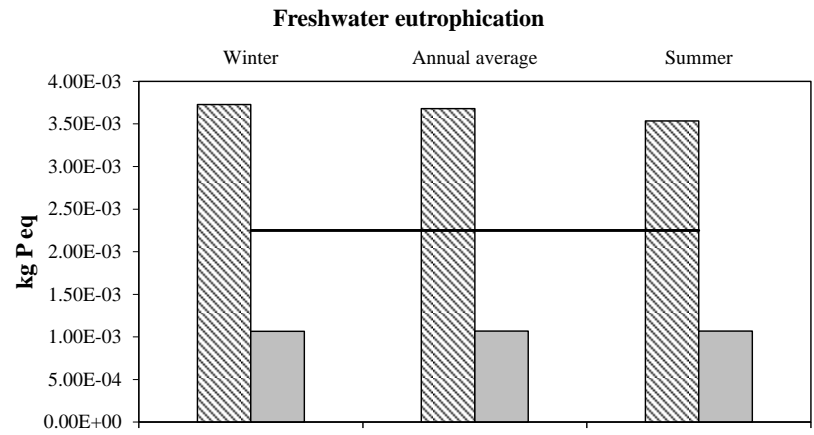

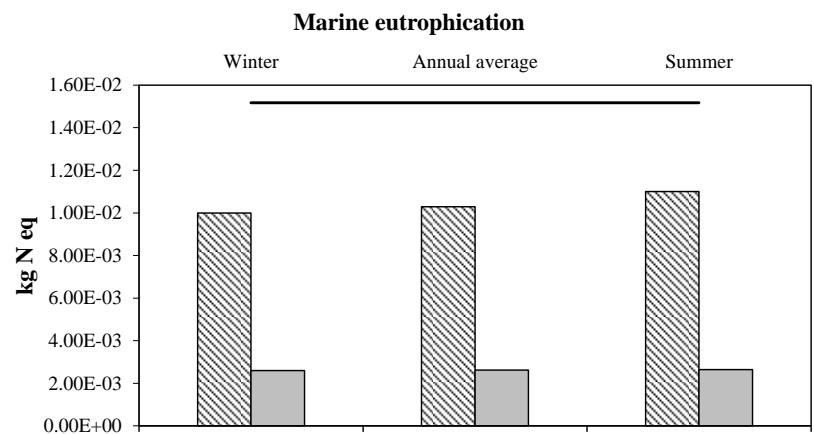
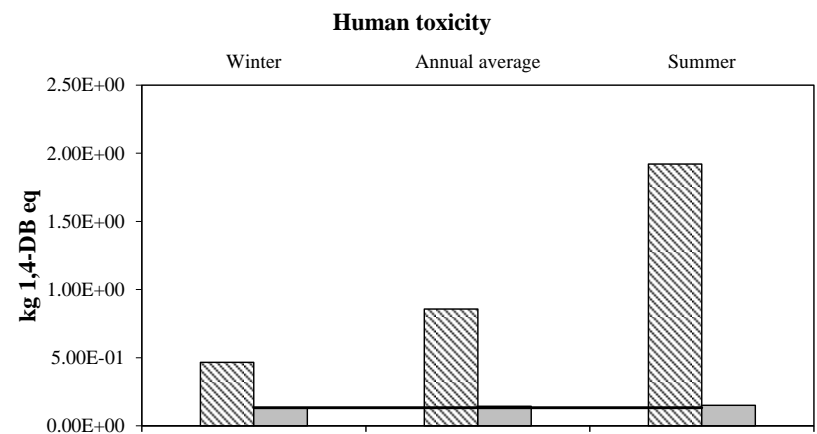
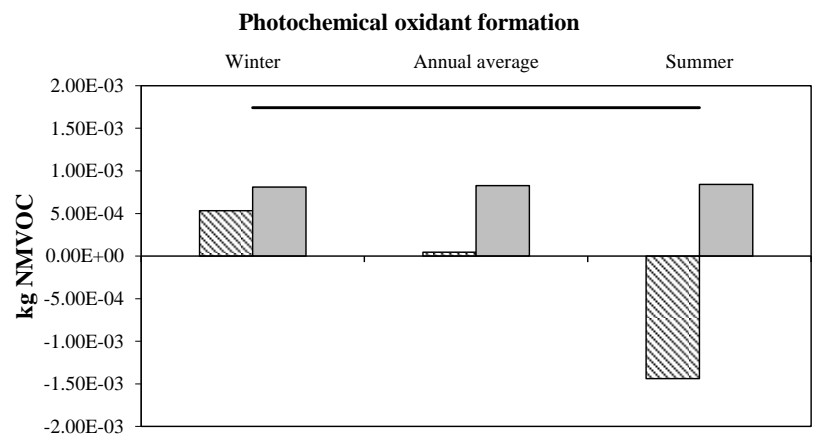
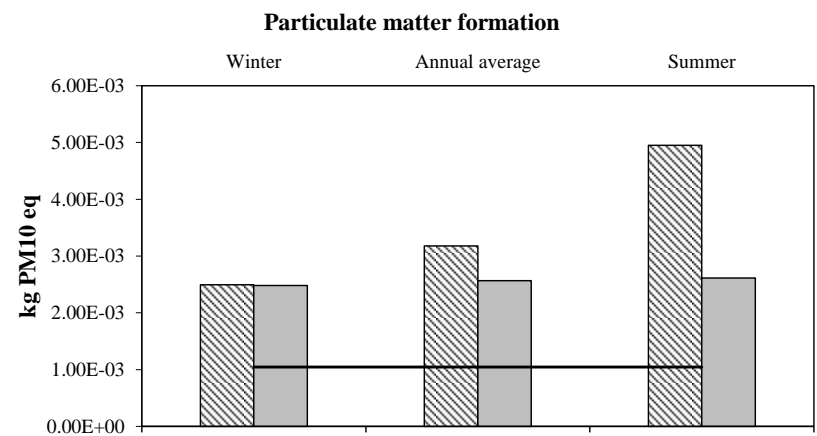

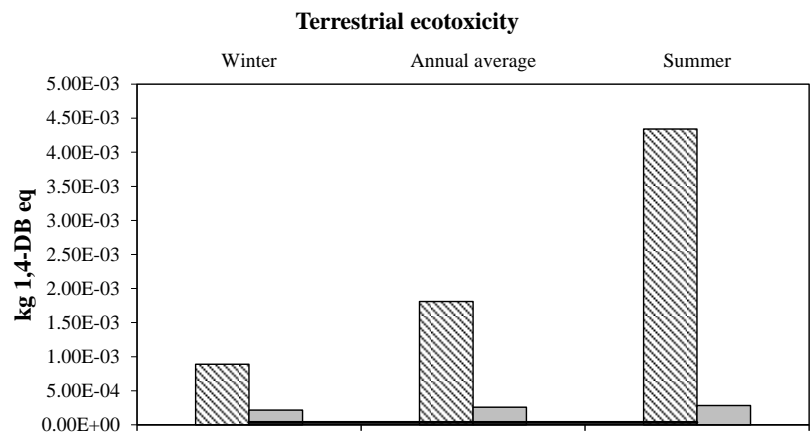
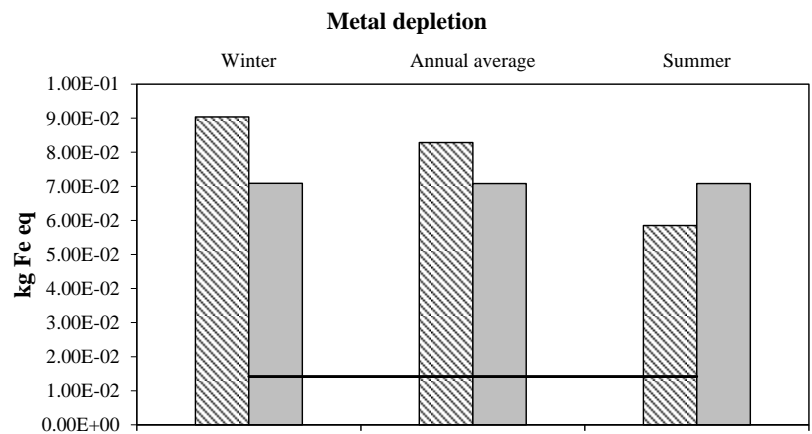
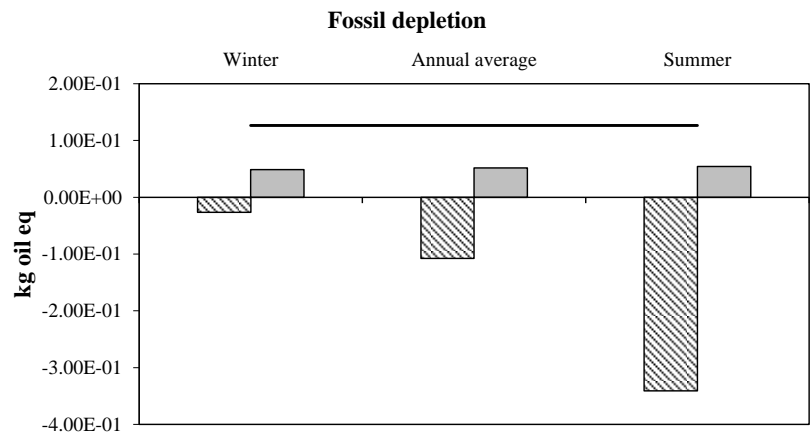

Scenario 1    Scenario 2    Scenario 3

**Figure 4.** Seasonal variation of the potential environmental impacts for the three scenarios: a) HRAP system for wastewater treatment where microalgal biomass is valorised for energy recovery (biogas production) (Scenario 1); b) HRAP system for wastewater treatment where microalgal biomass is reused for nutrients recovery (biofertilizer production) (Scenario 2); c) activated sludge system (Scenario 3). Values are referred to the functional unit (1 $m^3$ of water). Potential environmental impacts were calculated considering the microalgal biomass production achieved in summer and winter months (highest and lowest production, respectively).